# Buffalo Genome Projects: Current Situation and Future Perspective in Improving Breeding Programs

Ahmed M. Mousbah[1], Hesham M. Abdullah[1*,$], Waleed S. Mohammed[1], Ali M. El-Refy[1], Mohamed Helmy[2,*]

[1]Biotechnology Department, Faculty of Agriculture, Al-Azhar University, Cairo, 11651, Egypt

[2] Department of Computer Science, Lakehead University, ON, Canada

[$]Present address: Department of Plant Biology, Michigan State University, East Lansing, MI 48824, USA.

Email:

A.M.M: mousbah.biotech2012@gmail.com

H.M.A: abdull76@msu.edu

W.S.M: waleed.shaapan@azhar.edu.eg

A.M.E: elrefy67@azhar.edu.eg

M.H: Mohamed.helmy@bii.a-star.edu.sg

*Corresponding Authors:

**Hesham M. Abdullah, PhD**

Biotechnology Department, Faculty of Agriculture, Al-Azhar University, Cairo, 11651, Egypt

**Email:** abdull76@msu.edu

**Mohamed Helmy, PhD**

Mohamed.helmy@bii.a-star.edu.sg

## Abstract

Buffaloes are farm animals that contribute to food security by providing high-quality meat and milk. They can better tolerate the adverse effects of global climate change on their meat and milk production. Despite their advantages, buffaloes are heavily neglected animals with fewer studies compared to other farm animals; hence, the real potential of buffaloes has never been realized. The complete genome sequencing projects of buffaloes are essential to better understanding the buffalo's biology and production since they allow scientists to identify important genes and understand how the gene networks interact to determine the critical features of buffaloes. The genome projects are also valuable for gaining better knowledge of growth, development, maintenance, and determining factors associated with increased meat and milk production.

Furthermore, having access to a complete genome of high-quality and comprehensive annotations provides a powerful tool in breeding programs. The current review surveyed the publicly available buffalo genome projects and studied the impact of incorporating genomic selection into the buffalo breeding program. Our survey of the publicly available buffalo genome projects showed the promise of genomic selection in developing water buffalo science and technology for food security on a global scale.

## Background

The Food and Agricultural Organization (FAO) has described the Buffalo (Bubalus bubalis) as a strategic asset that is undervalued (1). Buffalos are called "Black Gold" because they contribute to food security and the economy as providers of milk, meat, hides, draught power and bone sources worldwide (1,86,91). Buffalo milk is unique and healthy because it is protein-rich, creamy in taste, and has lower cholesterol levels than cow's milk. Therefore, Buffalo's milk is appropriate for mozzarella cheese and yogurt manufacturing (2,3,87,88,89). Compared to cow's meat, buffalo meat contains less fat and cholesterol, and there is no religious prohibition against its consumption. As a result, it has many appealing characteristics for both consumers and breeders (4). Furthermore, buffaloes have higher disease resistance than cattle (5,88,90).

In the last few decades, buffaloes have attracted the attention of researchers worldwide. An accumulated amount of data has been generated on productivity and fertility in buffalo-related research institutions in many countries (5,91). Furthermore, many non-governmental organizations (NGOs) have scientific programs focusing on buffalo-related research, such as the International Buffalo Federation, the American Water Buffalo Association, and the Asian Buffalo Association (5–8, 91).

A significant aspect of buffalo-related genomic research is the genome sequencing projects of different buffalo species. The animal's genomic sequence is a crucial starting point for comparative genomic research that characterizes, analyzes, and protects the world's buffalo population and genetic diversity (9, 84,92). A well-annotated genome sequence also provides a foundation for modern systems biology approaches. Those approaches are utilized to optimize the genetics and metabolism of the animal, produce animal products in the lab (e.g., lab-grown meat), or produce plant-based products that are similar in properties to animal products (10).

For instance, the International Buffalo Genome Consortium (IBGC) gathered and drew the outlines for the buffalo genome project, which has three main phases. Phase one: Getting the entire genome sequence for the Mediterranean Buffalo. Phase two: Detection of single nucleotide polymorphism (SNP) and variation in domestic buffaloes. Phase three: comparative genomics, which includes sequencing the swamp buffalo genome, genome assembly using the river buffalo as a template, comparative genome analyses, and population genetics (11). On the other hand, a de novo assembled and thoroughly annotated water buffalo genome is a crucial requirement for researching this species' biology. Managing genetic diversity and utilizing modern breeding and genomic selection techniques are also essential (12,85,92). While large consortiums traditionally performed gene sequencing, recent technological advancements have brought whole-genome sequencing of non-model organisms within reach of a single laboratory (13).

In this review, we surveyed the publicly available buffalo genome projects worldwide, including the projects performed in China (14), Australia (15), India (16,17), Bangladesh (18), Italy (12), and Egypt (19). We technically evaluated these genome projects based on the sequencing system, assembly technique, and outcome. Furthermore, we provided a future perspective on the potential genomic research that used the outcome of these projects to advance the science and technology of water buffalo for food security on a global scale.

## The Origin, Domestication, and Classification of the Buffaloes

According to a recent estimate, there are 224.4 million buffaloes globally, 219 million of which (97.58%) are found in Asia (83,17). The rest is spread across some 129 countries worldwide, mainly tropical and sub-tropical countries (12) (Figure 1).

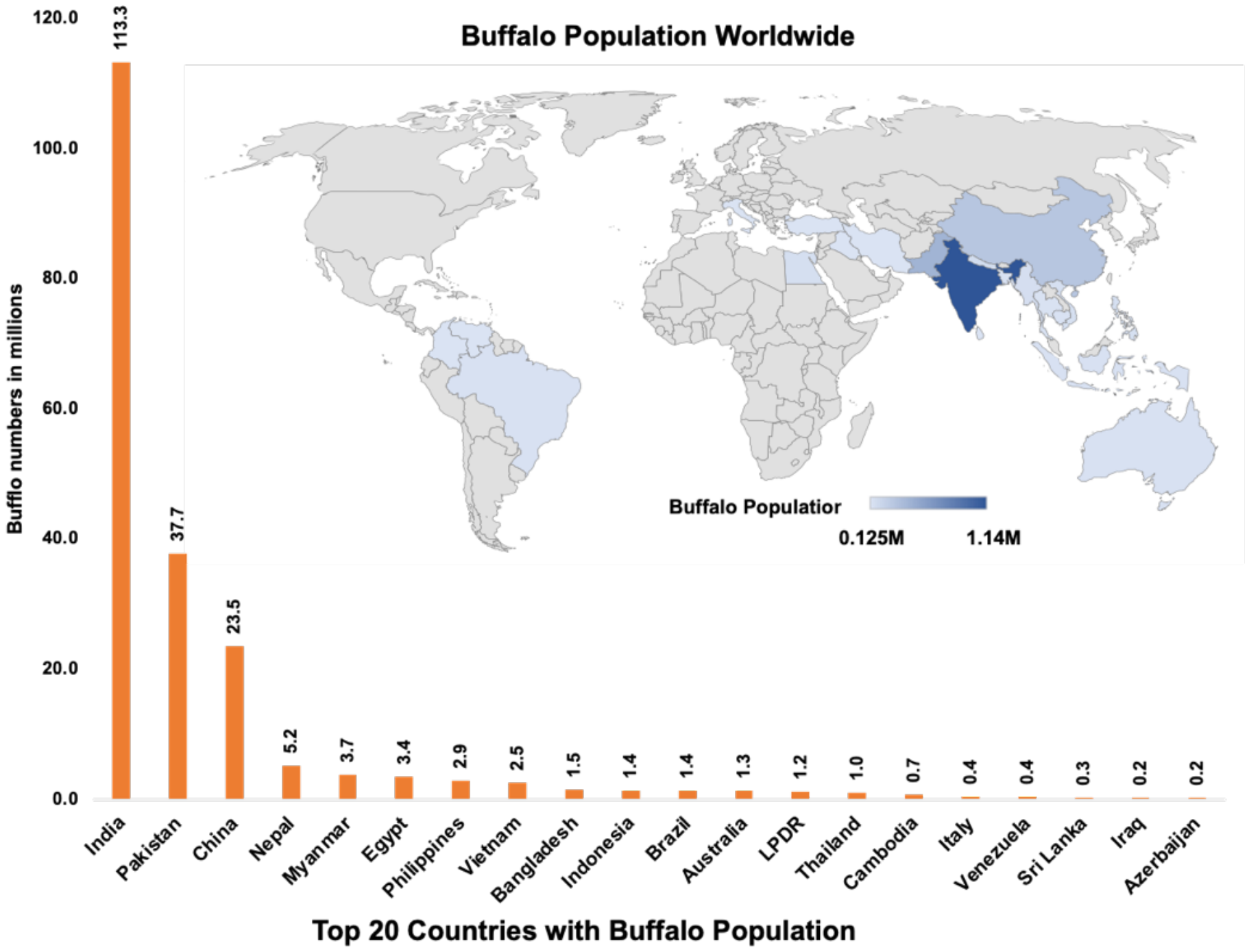

**Figure 1**

Asiatic water buffalo have historically been divided into river and swamp buffalo based on physical, behavioral, geographic, and genetic variables. The two types were also found to have a varied number of chromosomes (river 2n = 50, swamp 2n = 48) (20,21, 92).  Asian water buffalos are sometimes classified as belonging to two separate species (swamp buffalo as *Bubalus carabenesis* and river buffalo as Bubalus bubalis) or two different subspecies (river buffalo as *Bubalus bubalis* and *Bubalus bubalis* for river buffalo) (21–23).  It is important to note that the water buffalo, which originates in East Asia, is thought to account for more than half of the world's buffalo population (6). These species are thought to have originated in mainland Southeast Asia before migration to China and the Indian subcontinent, respectively, in the north and west (24) (Figure 2).

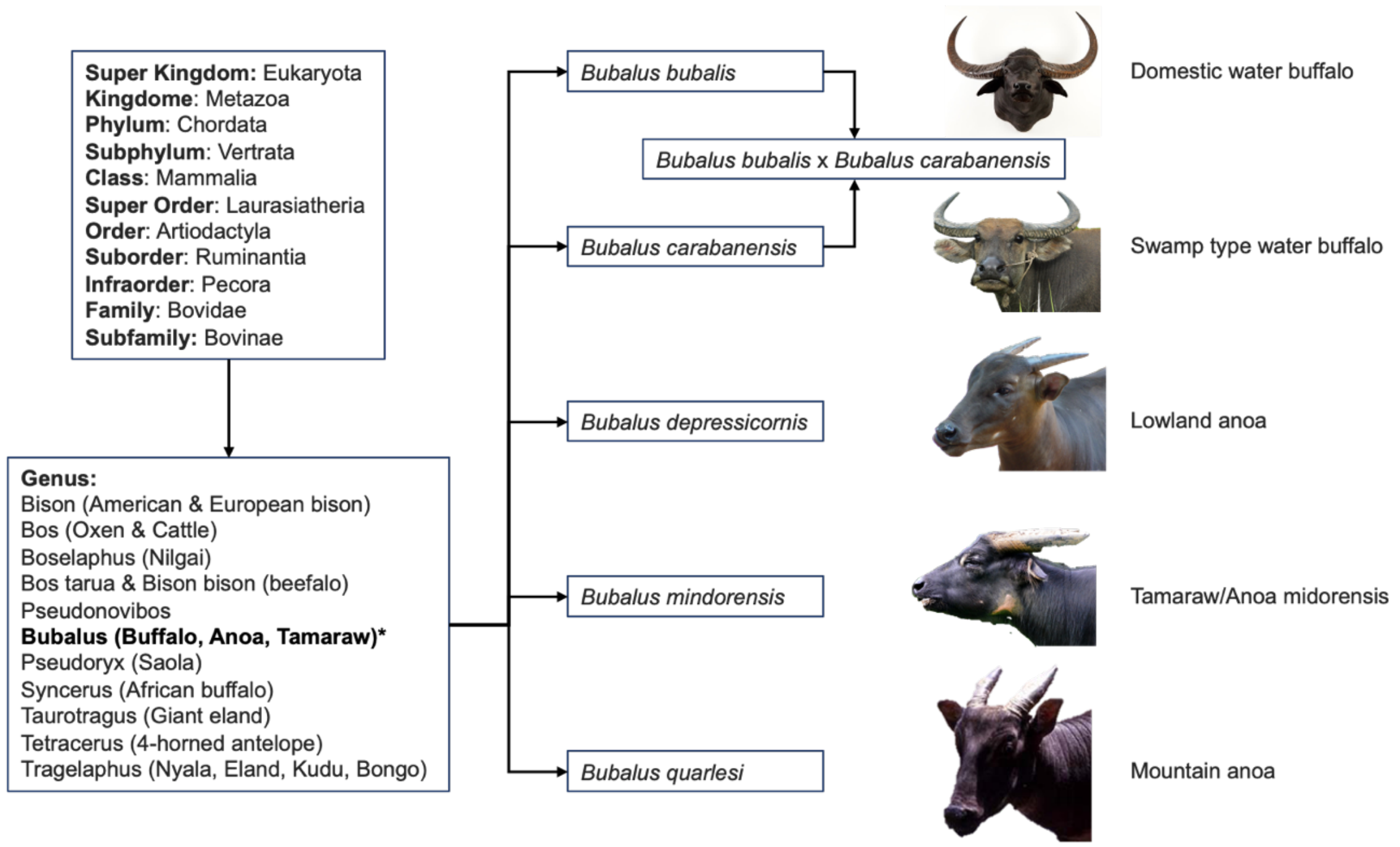

**Figure 2**

It is believed that the buffalo was tamed in the Indus valley of the Indian subcontinent between 4000 and 5000 years ago (25). The domestication of the river buffalo took place in the Indian subcontinent around 6.3 kyr BP, according to the whole mitogenomes reported in (26). Comparatively, the domestication of the swamp buffalo began in China between 3 and 7 kyr BP (26).

Many researchers determined that the swamp and river types have diverged between 10 kyr and 1.7 Myr BP long before domestication (27–29). The divergence time was calculated to be between 900 and 860 kyr BP and then evolved in different geographical areas, according to a recent study that analyzed whole-genome data (14). The period determined by the MCMC tree was around 3.5 million years ago, but the timing anticipated by the distribution of Ks was roughly 1.1 million years ago (14). The Bovini tribe, including domesticated and wild buffalo and cattle species, began to separate more than four million years ago. The subfamily of the Bovinae (Family Bovidae, order Artiodactyla, class Mammalia, phylum Chrodata, Kingdom Animalia) is divided into the tribes of Boselaphini (nilgai, four-horned antelope), Tragelaphini or Strepsocerotini (spiral-horned antelopes), and large-sized Bovini (30). Bovini is split into the genera Syncerina (African buffalo), Bovina (Bos and Bison species) and Bubalina (Bubalus species, water buffalo, arni, tamarao, and anoa) (31). Molecular evidence points to a distinction between non-buffalo (Bos, bison) and buffalo (Bubalus, syncerus) (30). Bison (bison and wisent) and Bos are considered subgenera of the family Bovinae. In alternative classifications, the Bos species are divided into the Bos (taurine and indicine cattle), Bibos (gaur, banteng, and kouprey), and subgenera poephagus (yak), Bibos (gaur, banteng, and kouprey), (Figure 2). However, this evidence does not support the hypothesis that yak and bison species are clustered together (32–35).

## Buffalo Standard Karyotype

Cattle and other members of the Bovidae family share morphological, physiological, and genetic similarities with water buffaloes. However, the numbers of chromosomes in these species vary (50, 48, and 60 chromosomes) in river buffalos, swamp buffalos, and cattle, respectively. C-banding and cytogenetic studies revealed the difference in chromosome numbers between the two buffalo subspecies and the arm-by-arm matching of water buffalo and cattle chromosomes (36,37). According to cytogenetic and C-banding research, the two buffalo subspecies differ in terms of chromosome numbers and the chromosomes of cattle and water buffalo match one another arm by the arm (36,37). According to the standard karyotype of the river buffalo and its homology with cattle, Researchers should focus on river buffalo chromosomes (1 p, 2p, 4p, and 5p) that are homologous and correspond to (25, 23, 28, and 29) chromosomes of cattle, respectively (38). The sex chromosomes are found in the remaining 20 pairs, which are acrocentric. After the advances in animal genetic mapping studies, some modifications in chromosome numbers were introduced (39), and homologous chromosomes among cattle and buffalo karyotypes were corrected, as shown in Table 1. The Radiation Hybrid (RH) mapping approach has been widely employed for producing medium-to-high-resolution maps. For creating medium-to-high-resolution maps, the Radiation Hybrid (RH) mapping method has been frequently used (40). Panels of RH maps could be accessible for many livestock species. The RH maps for individual river buffalo chromosomes are created using the RH panel. (40). The RH panel generates the RH maps for each river buffalo chromosome (40). River buffalo's first-generation whole genome RH map was created using cattle genomes as reference genome markers, which is a valuable source of markers for mapping the buffalo genome (41). The above-mentioned genome project included 2,621 loci obtained from cattle, covering all river buffalo chromosomes. (41). The arrangement of markers inside linkage groups in the RH map of the river buffalo genome was quite similar to that of the cow genome (42).

## Bovine Genome Projects Worldwide

Animal breeding operations in farm animals have been revolutionized thanks to genome technology. Sheep (Ovis aries) (43), Pig (Sus scrofa) (44), Horse (Equus caballus) (45), Wade, Goat (Capra hircus) (46) and Chicken (Gallus gallus) (47) all have genome projects that have offered information and foundations to maximize their capacity for productivity, reproductive effectiveness, and disease resistance (Table 2).

Knowing the genomic sequences of cattle such as cows and bulls is important because it provides resources for researching the evolution of mammals and speeds up livestock genetic breakthroughs for producing high-value milk, meat, and hide (84). Therefore, the bovine subfamily has been subjected to genome sequencing projects over the last decades, and many genomes from members of this subfamily have become available in public biological databases. In this section, we highlight the importance of genome sequencing projects for farm animals by reviewing some of the genome projects of some buffalo-related animal species that are published or still in progress (15, 43, 44, 45, 46, 47, 48). The information available from related species such as cattle, cows, and pythons (also called American buffalos) would guide the research on the buffalo genomes as tools to identify candidate genes that encode critical traits utilized in buffalo breeding programs (48-50, 51).

Among the cattle genomes in the subfamily Bovine, the whole genomes of *Bos taurus* and domestic yak (*Bos grunniens*) have been fully sequenced in the years 2009 and 2012, respectively (48–50). Those cattle genome projects would contribute to understanding the cattle's biology and accelerate the breeding programs to improve milk and meat production by exploring the mechanisms underlying the regulation of milk and meat production traits and identifying candidate genes that encode key traits utilized in buffalo breeding programs**.**

Later in 2015, the American Bison (Bison bison) reference genome of 2.82 GBP was assembled and became publicly available (51). North American bison's first genome sequence analysis has provided helpful information about the genetic background, taxonomy, and inheritance of important genetic traits. Knowing this information has helped us learn more about how these species survived for years and years, regardless of environmental changes. The bison species could advance their abilities to manage and thrive during the recovery from population destruction that occurred over hundreds of years. As an output of the genomic studies, genomic variants were identified in bison by aligning with the genomes references from domestic cattle.

The annotation of the identified variants has provided evidence for the genetic architecture of the variations and conservations that existed in the genomes between the modern bison and the historic bison (51).

After being compared to genome references from domestic cattle, the bison reference genome helped uncover new genetic variations, such as insertion-deletion mutations (INDELs) and single nucleotide variants (SNVs) at the genomic level. This comparative research revealed 30 million novel variations (both SNVs and INDELs combined) between bison and domestic cattle. (51). The matching genes and functions of these discovered variants were identified to explain the genomic variations between domestic cattle and bison (51).

## The African Buffalo (*Syncerus caffer*) Genome

African countries such as Ethiopia, Tanzania, Namibia, Mozambique, South Africa, Kenya, Somalia, Zambia, Zimbabwe, and Botswana are home to the African buffalo, a wild of buffalo that originated in central Africa. The population of the African buffalo was estimated at 900K (52). However, there has recently been reduced (36). The African buffalo genome was sequenced and mapped using two separate software packages to find novel SNPs in the genome on an abroad scale, and fluorescence-based genotyping was used to compare the SNP validation rates (53). The Bos taurus UMD3.1 genome sequence, generated by the University of Maryland's Centre for Bioinformatics and Computational Biology (CBCB), was chosen as the reference genome in this research study because of the lack of an assembled African buffalo genome. The African buffalo genome (Syncerus caffer, 2n=52) was assembled as a 2.68 Gb contig with an N50 of 43 kbp and a Scaffold N50 of 2.3 Mbp (53). The genome scan discovered a total of 19,765 genes, with 97.60 % having satisfactory annotations. Additionally, they found a number of extended predicted genes linked to various GO terms, including G-coupled proteins and olfactory receptors (53).

In March 2018, Northwestern Polytechnical University in China submitted to the Gene Bank (GCA_006408785.1) a genome sequencing and assembly of an African buffalo using Illumina HiSeq (162X) (54). More recently, in March 2020, the University of Edinburgh in the United Kingdom submitted the first high-quality genome assembly of the African buffalo (Syncerus caffer) based on a combination of PacBio (60X), Illumina (75X) and Hi-C libraries for scaffolding to the Gene Bank (GCA_902825105.1). It is essential to have an accurate reference genome for the African buffalo to understand the genomic variation across the Syncerus caffer sub-species.

## The Water (Asiatic) Buffalo (*Bubalus bubalis*)

In recent years, genome projects of water buffalo have intensively increased, reflecting its importance to understanding the biology of different species in applying genomic selection to promote the genetic characteristics of livestock. The information from the publicly available genome projects of water buffalos worldwide is summarized in Table 3.

### *a) Chinese Water Buffalo Genome Project*

The Chinese water buffalo genome project was targeted two buffalo species: river and swamp buffaloes. A total of 98 river buffaloes and 132 swamp buffaloes from Asia and Europe were sequenced and assembled. The authors compared the genomes of male and female buffaloes; a male from the Fuzhong swamp buffalo and a female from the Murrah river buffalo were used for sequencing. Using Illumina High throughput/resolution chromosome conformation capture sequencing (Hi-C), the produced contigs from PacBio assemblies were scaffolded and clustered.

The contigs were grouped together by Hi-C sequencing into 24 chromosomes, which covered 97.5 % of the swamp buffalo's genome and 25 chromosomes, which covered 96.5% of the river buffalo's genome. According to the final assembly and nucleotide accuracy evaluation, the error rate in swamp buffaloes was 9.22×10−6, and in river buffaloes, 2.13 × 10−5 (14, 84). Luo and his co-authors analyzed selective-sweep methods, different demographics, and population structure (14) to understand better the genetic and physiological diversity and the evolution of these important animal species. The authors found swamp buffalo-specific selection for brain development and cognition-related genes, which may help explain the genetic basis for the obedience that enables these animals to serve as long-term collaborators in the development of the rice-paddy agro-ecosystem (14). Additionally, particular genes associated with fertility, milk production, and body size specific to river buffalo were discovered from that study, which benefited in the full process of choosing and breeding together with genome-wide association analysis (GWAS) (14,84).

***b) Australian Buffalo Genome Project***

Low and co-workers in 2019 published the Australian water buffalo (*Bubalus bubalis*) genome with a chromosome-level assembly using single-molecule sequencing and chromatin conformation capture (Hi-C) approach. Sequence data were generated from ~75x PacBio Sequel long-reads, ~24x Chicago reads, ~58x Hi-C reads, and ~82x Illumina paired-end reads (15). The assembly was performed using the FALCON-Unzip assembler (55), which produced 953 contigs with a N50 of 18.8 Mb and a total size of 2.65 Gb. Linkage disequilibrium maps were generated for each of the 457 contig junctions in the critical 29 scaffolds representing the 25 buffalo chromosomes to assess the assembly. They identified 29,244 genes count, 58,204 coding sequences counted, 238,481 introns counted, 269,697 exons count, and 71,537 transcripts count, with a mean number of exons per transcript of 2.54 (15). The project makes it easier to annotate gene clusters that are difficult to put together, such as the major histocompatibility complex (MHC). The MHC is crucial for disease resistance because it plays a crucial role in triggering immunological responses (15).

### *c) Indian Buffalo Genome Project*

Over the last decade, India has completed two genome projects (16, 17). The first project presents the initial assembly of a single female Murrah buffalo that was sequenced using Illumina technology at a depth of 17-19X, utilizing as a reference the cattle genome (Btau 4.0 assembly). The results were deposited in GenBank under the accession #PRJNA33659 ( 16). The coverage of the buffalo assembly represents about 91% ~ 95% compared to the reference cattle genome. The buffalo assembly's coverage ranges from about 91% to 95% compared to the reference cattle genome. The raw reads were assembled into 185,150 contigs, with the biggest contig at 663 Kb and the median contig length being 2.3 Kb. The mitochondrial genome was assembled into a single separate contig ( 16 ). The information in this assembly are used to learn more about the genetic variation that set buffalo apart from cattle as well as adaptive traits from genomic perspectives. . The second project provides the first improved haplotype phased river buffalo genome assembly (Murrah Indian breed) (17). This genome project is designed based on the sequencing of a mother-father-progeny triple. The Illumina and PacBio long-read platforms generated 274 Gb of paired-end reads from parental DNA samples and 802 Gb optical mapping data from the progeny DNA samples, respectively. Each haplotype was scaffolded based on the FALCON assembly to create a reference-quality assembly of 2.63 Gb and 2.64 Gb for sire and dam haplotypes with only 59 and 64 scaffolds, and an N50 of 81.98 Mb and 83.23 Mb, respectively. The chromosomal level assembly was constructed and organized using RaGOO, and it included 25 scaffolds with N50 values of 118.51 Mb (dam haplotype) and 117.48 Mb (sire haplotype) (17). The outcome of this project is an essential resource for the in-depth exploration of mechanisms underlying the regulation of milk and meat production traits to improve buffalo breeding programs (17).

### *d) Bangladeshi Buffalo Genome Project*

The first river buffalo genome from Bangladesh was released in March 2012 due to the scientific collaboration between a Bangladeshi company (namely Lal Teer Livestock) and the Beijing Genomics Institute (BGI) in China. The two organizations have jointly sequenced the complete genome of the water buffalo. The project helped improve breeding programs for buffalo strains with better milk and meat productivity. The combined efforts led to the discovery of 24,613 protein-coding genes and a high-quality water buffalo genome with a genome size of roughly 2.77 Gb (18,84).

The project was conducted to identify the buffalo genome components and genes of industrial value. The researchers found that the water buffalo genome had 1,418 Mbp of repetitive DNA, which made up 51.19 % of the entire genome; additionally, a total of 38,483 tRNAs, 23,310 miRNAs, 1,758 snRNAs, and 867 rRNAs were annotated (18). The researchers also found that 159 gene families were significantly larger in water buffalos than in other mammals. Those genes were functionally annotated to gene families related to buffaloes' adaptation to the surrounding environments, signaling receptor activity and olfactory receptor activity, ATP-binding, G-protein-coupled receptor signaling pathway, and transmembrane transport, signal transduction pathway, metabolic pathway, immune system functional pathway (18,84).

### *e) Italian Buffalo Genome Project*

The International Water Buffalo Genomic Consortium carried out *de novo* assembly of an Italian Mediterranean river buffalo (12). This is considered the first river buffalo genome in Italy *had de novo* sequence assembly and annotation. The project used (MaSuRCA) assembler version 1.8.3 (56). Two parallel assembly run with different parameters and generated 2.83 Gb, which was made up of 366 983 scaffolds with N50 values of 1.41 Mb for the scaffolds and 21 398 bp for the contig. They identified 27,837 genes and pseudogenes, 21,711 protein-coding, 230 non-coding, 3,823 pseudogenes and 41,486 mRNA (12). Compared to the earlier alignment of low-coverage to the bovine genome sequence, this project was regarded the first de novo sequence assembly and annotation of the river buffalo genome (12). ***f) Egyptian Water Buffalo Genome Project***

The Egyptian Water Buffalo Genome Consortium released the preliminary results of the domestic Egyptian buffalo breeds completed in 2013. The high-quality annotated genome version was finished in early 2015 (19).

SOLiD Sequencing technique was used for the sequencing. The scaffold-level assembly was finished using a reference-based assembly to improve the assembly quality. The Bovine genome (Bos taurus UMD 3.1.1) was utilized by the authors as a reference genome. Using trimmed reads (Q15), 447,289 contigs with a total size of 3.005 Gb were constructed. The largest contig was 148 Kb, while the N50 for contigs was 14.568 Kb. (19). The expected outcome of this genome project is the generation of genomic datasets on buffalo genetics, which will allow the breeding programs to identify potential targets to increase Egyptian buffalo's ability to produce more meat and milk (19) (Table 3).

## Buffalo Transcriptome and SNP Chip

Transcriptomics is an existing technology for measuring gene expression and regulation (transcriptome) and interpreting the genome's functional elements. A better understanding of the buffalo transcriptome enables researchers to identify potential genetic variants affecting important economic traits and accelerate the genetic improvement in water buffalo breeding programs by the buffalo breeding industry (3,57–60). The best available and specific tool for the genomic characterization of domestic buffaloes is the commercial buffalo SNP chip array, Axiom Buffalo Genotyping array 90 (Affymetrix) (9) which updated by. Nascimento and his colleagues in 2021(93). Before the Axiom Buffalo Genotyping Array, several attempts were made to detect water buffalo genomic variability using the bovine SNP chip (Illumina Bovine SNP 50 Bead chip and 777k Illumina Infinium BovineHD Beed chip) with cattle-specific arrays. (42,61–64). The availability of a commercial buffalo-specific SNP chip array allowed studying several aspects of buffalo's biology and production, including milk production (2,3,9,63,65,66,88,89,94,95), reproductive traits (82, 88), animal management and breeding (67), population diversity (21,59,68, 92,94), genetic disorders (69), heat tolerance (58), and data analysis application (70). In addition to the previous research that used the commercial buffalo SNP chip, other buffalo transcriptome studies were reported, utilizing different approaches. For example, 52,979,055 high-quality RNA-Seq reads were assembled de novo to provide 86,017 unigenes, with 62,337 successfully annotated (57). Additionally, a thorough an atlas of gene expression across 220 tissue and cell samples was released in another study (60), and it was used to annotate the newly assembled buffalo genome (15).

## Discussion

Contrary to the cattle genome, which has been sequenced and well-defined (the bovine genome sequencing and analysis) (49,71), little has been done to uncover the water buffalo genome structure and annotation. Therefore, in 2013, the International Buffalo Genome Consortium (IBGC) gathered and created the outlines for the buffalo genome project planned in three phases.

Phase 1: The Complete Sequencing of the Mediterranean Buffalo. This step includes de novo sequencing, genome assembly, genome annotation, and transcriptomic analysis.

Phase 2: SNP and variation detection in domestic buffaloes, including SNP discovery and the SNP genotyping panel setup.

Phase 3: Comparative genomics, which includes sequencing the Swamp Buffalo genome, genome assembly using the genome of the river buffalo as a reference genome, comparative genome analyses, and population genetics (11). The buffalo genome projects worldwide started after the International Buffalo Genome Consortium (IBGC) gathered to reference the buffalo genome.

***Buffalo Genome Projects Assessment***

Our survey and assessment show that both the Chinese and the Australian buffalo genome projects provide the best outcomes. The two projects used cutting-edge sequencing techniques, and both of them provided genome assembly at the chromosome level. However, the other buffalo genome projects presented herein used less advanced sequencing systems and provided genome assembly at the scaffolds and contigs level. The number of scaffolds and contigs is essential to judge the accuracy of the project's outcomes; the smaller the number, the better the outcome. Herein, we highlight the advantages and main features of the reviewed genome projects.

a) The Chinese buffalo genome project is the most extensive buffalo genome-sequencing project ever completed. The researchers in that project sequenced and analyzed the genomes of 230 geographically disparate buffaloes (98 river buffaloes and 132 swamp buffaloes) from various regions of Europe and Asia (14). This effort resulted in buffaloes reference genomes for river and swamp buffaloes and extensive SNP datasets, providing long-needed resources for buffalo genomic research (14).

b) The Australian buffalo genome assembly produced the species' current reference genome (15, 60). That reference genome, which has a 2.66 Gb overall length and 509 scaffolds with an N50 of 117.2 Mb, is partly phased. As a result, it has a high degree of sequence continuity and a low number of gaps (only 383). Through the use of long-read PacBio sequencing, the project covers numerous regions that were absent from the same genome’s Illumina-based sequence. Long-read sequencing allows for the incorporation of huge repeat families like LINE L1 and BovB, which are challenging to assemble using short-read-based techniques (60) correctly. The evolution of these elements cannot be explored without this information. These elements, for instance, vary between species and may affect how genes are expressed (for instance, LINE L1 and BovB make up 16% of the B. bubalis genome) (60).

Long-read sequencing, combined with chromatin conformation capture technology, is currently one of the most effective methods for generating high-quality genome assemblies without needing a reference genome. One of the best ways for creating high-quality genome assemblies without a reference genome is now long-read sequencing combined with chromatin conformation capture technologies. . The buffalo genome project in Australia produced over 21 billion raw sequence reads, which were then mapped to 18,730 distinct genes (15). These datasets helped the Australian buffalo genome assembly to annotate transcribed sequences (15).

The Australian project created a thorough atlas of gene expression, including cell samples and 220 tissue obtained from 10 river buffaloes of three distinct breeds (Bhadawari, Mediterranean, and Pandharpuri) (15 and 60). These datasets supported the annotation of transcribed sequences in the Australian buffalo genome assembly (15).

c) There are two buffalo genome projects in India. The first was published in 2011 and had sloppy results. In the Indian buffalo genome, the project only found a few SNPs and INDELs (16). Recent publications include the first haplotype-phased reference-quality genome assembly of the Murrah river buffalo breed from India (17). This improved haplotype phased genome assembly provides great resources for understanding the genetic mechanisms behind buffalo milk production and reproduction features (17).

d) The sequencing and assembly of the Bangladesh river buffalo genome was accomplished by Bangladesh genome projects (18). The genome was 2.77 Gb in size and had scaffold and coting N50 of 6.9 Mbp and 25 kb, respectively (18). The assembled genomes were compared to the Mediterranean water buffalo and the African buffalo publicly accessible genomes. To assess genomic completeness, the researchers used two alternative approaches. To begin, the EST/mRNA sequences were retrieved from the NCBI, and they used BLAST to align them with the completed genome as per (72). Using this method, they aligned about 98.15 % of the data. Second, the assembled genome had 94.3 % coverage of the essential genes thanks to benchmarking against universal single-copy orthologs (BUSCO 2.0) (73). The water buffalo genome (39.578) has more CpG island sequences than the cattle genome (74). Additionally, 382 candidate genes with sites for positive selection were found in the study, which could help water buffalo acquire climate tolerance in a diverse environment (18).

After releasing the Australian buffalo genome, the Axiom Buffalo Genotyping Array has recently been remapped (15). e) The Italian project initiative can be viewed as the most prosperous buffalo genome project from an economic point of view, it resulted in a patent for the Axiom Buffalo Genotyping 90K array (Affymetrix) as a water buffalo-specific genotyping tool (9). It was used in several successful projects worldwide to enhance buffalo productivity and fertility (2,3,9,21,58,59,63,65–70). However, the Italian project with the final assembly containing 2.84 Gb spread across 366,983 scaffolds and a contig N50 of 22 kb, was highly fragmented (15).

f) The Egyptian water buffalo genome project utilized short-read sequencing using the SOLiD platform, which resulted in a total of 3.4 billion reads with a 229.24 Gb. The assembly of longer contigs was challenging due to the SOLiD platform's short read-length output. Furthermore, the genome's repeats create uncertainty that cannot be resolved using short reads/contigs in the assembled areas (19). The recent buffalo assembly reveals that modern long-read sequencing tools offer a huge advantage in building better scaffolds (14).

## Future Perspectives

Based on the previous review of all the buffalo genome projects worldwide, we can see that the buffalo genome is understudied, despite its importance for food security and the ecosystems in its lives. Further studies should focus on the buffalo genomic structural and functional annotations. This will allow the research community to investigate the mechanisms regulating the feature that lead to the production of milk and meat and disease resistance mechanisms. Moreover, we suggest the following:

1. More large-scale buffalo genome sequencing projects should be carried on to better

understand buffalo biology on the molecular level and to study variations and SNPs for improving breeding strategies

2. Creating a centralized buffalo genome resources center to aid buffalo genomics research by providing a database of available buffalo genomes, genome annotations, and data mining tools. The resources center should contain all the updates from buffalo genome projects worldwide and the updates from international buffalo meetings, such as the International Buffalo Federation, the American Water Buffalo Association, and the Asian Buffalo Association. In addition, links related to buffalo-specific tools such as the BuffSatDb Buffalo Microsatellite Database (75) and miRBL, the database for microRNA of water buffalo (76),

3- More studies are required to characterize the buffalo transcriptome to provide insights into the mechanisms regulating the production, climate change tolerance, and disease resistance traits in buffalo. This will assist in locating prospective candidate genes that could be used for later functional characterization of buffalo milk and meat production and could be applied to buffalo breeding programs. The accessibility of the entire transcriptome data for buffalo would also be a resource for improving the genome annotation through inference of the exon-intron boundaries and, therefore, better identification of the functional regions in the genome.

4. There is a growing awareness concerning buffaloes as food security animals that could survive climate change and global warming challenges. Climate change has seriously endangered the world's livestock production systems' long-term viability. Water buffaloes are more adaptable to hot and humid climates than other domestic species because of their morphological, anatomical, and behavioral characteristics. Buffalo skin has a thick epidermis rich in melanin pigments, which give the animal's skin its characteristic black color and protect it from the harmful effects of UV rays (77). In addition, buffalo have well-developed sebaceous glands that secrete sebum, a fatty substance that acts as a lubricant. The sebum layer melts and gets glossier in hot weather to deflect the sun's rays and reduce the animal's high exterior heat burden (78). The strategies for decreasing the influence of heat stress on the productivity and reproduction of water buffalo, along with better management and identification of SNPs responsible for variation in animal sensibility or tolerance to thermal stress, will enable the selection of water buffalo with heat tolerance traits.

5. Restructure breeding programs to capitalize on genomic selection more effectively. Implementing genomic selection will lower the cost of the breeding program, boost the rate of genetic gain within the breeding nucleus, and increase selection accuracy. Thanks to the genomic selection, the generation intervals in the male-to-male selection pathway could be reduced from 9.5 to 3.3 years. The male-to-male selection route could lower the generation intervals from 9.5 to 3.3 years. Compared to progeny testing programs, it can also result in a two-fold increase in selection response and an 88 percent reduction in proven bull costs (79). Two promising studies provided examples of using genomic selection in buffalo and their future application.

1. Pakistan (Moaeen-ud-Din and Bilal (2016), (79) illustrated detailed examples to compare traditional progeny testing vs genomic selection and evaluated the Selection, Intensity, Accuracy and Generation Interval, and cost, in addition to designing a promising Genomic Selection Program for buffaloes. 2. Italy, (Cesarani et al., 2021, (85).) used genomic models to predict specific traits in the Italian Buffalo and evaluated the promising results of genomic selection in breeding programs.

## Conclusions

The present survey of the water buffalo genome projects offers an opportunity to assess the completeness of the available genomic water buffalo information. It shows that the available genome sequences, structural annotations, and functional annotations are far from what is required to study and understand such important animals. Therefore, improved genomic information could aid in the future discovery of genes linked to complex features involved in meat and production, as well as, disease resistance and susceptibility. This will enable researchers, animal breeders and policymakers to effectively improve economic features genetically through traditional breeding programs using genomic breeding for better selection in a shorter time and at a reduced cost.

## Declarations

### Ethics approval and consent to participate

Not applicable

### Consent for publication

Not applicable

### Availability of data and materials

Data sharing is not applicable to this article as no datasets were generated or analyzed during the current study.

### Competing interests

The authors declare that they have no competing interests.

### Funding

Not applicable

### Authors' contributions

All the authors conceived the study and oversaw its design and coordination. All the authors contributed to data collection and analysis, discussing the results and representing the key findings in figures and tables. HMA, WSM, and AME contributed to the experimental design and supervision of the project. All authors have been involved in drafting the review article and/or revising it critically for important intellectual content. All authors have read and approved the final manuscript.

## Footnotes

### Figure Legends

**Figure 1.** Distribution of the buffalo population worldwide. The map shows the countries with a buffalo population of more than 100K. The bar chart shows the top 20 countries with buffalo populations. The data presented herein were obtained from (6).

**Figure 2.** Taxonomy classification of the water buffalo into the animal kingdom. The diagram was based on data obtained from NCBI Taxonomy (80). As discussed in the main text* The classification of the genus *Bubalus* is debatable, so we did not designate the different buffalos as species or subspecies.

| **Table 1. Homologous chromosomes among cattle and river buffalo karyotypes** | | | | | | | | | | | |
|---|---|---|---|---|---|---|---|---|---|---|---|
| **River buffalo** | 1q | 1p | 2q | 2p | 3q | 3p | 4q | 4p | 5q | 5p | 6 |
| **Cattle** | 1 | 27 | 2 | 23 | 8 | 19 | 5 | 28 | 16 | 29 | 3 |
| **River buffalo** | 7 | 8 | 9 | 10 | 11 | 12 | 13 | 14 | 15 | 16 | 17 |
| **Cattle** | 6 | 4 | 7 | 9 | 10 | 11 | 12 | 13 | 14 | 15 | 17 |
| **River buffalo** | 18 | 19 | 20 | 21 | 22 | 23 | 24 | X | Y | | |
| **Cattle** | 18 | 20 | 21 | 22 | 24 | 26 | 25 | X | Y | | |
| **Data is adapted from (39).** | | | | | | | | | | | |

**Table 2.** Characteristics of assembled domestic animal genomes

| | Sequencing Strategy (Fold coverage) | Genome length | Protein coding genes | Release year | Reference |
|---|---|---|---|---|---|
| **Buffalo (*Bubalus bubalis*)** | WGS, PacBio (69.0x) | 2.66Gbp | 29,244 | 2018 | (15) |
| **Cow (*Bos taurus*)** | Whole-genome shotgun/BAC and other clones (7.1×) | 2.91Gbp | 19,241 | 2009 | (48) |
| **Sheep (*Ovis aries*)** | Whole-genome shotgun (3×) | 2.26Gbs | NA | 2008 | (43) |
| **Pig (*Sus scrofa*)** | Whole-genome shotgun (0.66×) and Minimal tile path BAC by BAC (6×) | 2.47Gb | 621 | 2005, 2009 | (44) |
| **Horse (*Equus caballus*)** | Whole-genome shotgun/BAC and other clones (6.8×) | 2.47Gb | 15,355 | 2009 | (45) |
| **Goat (*Capra hircus*)** | Single-molecule sequencing and chromatin conformation | NA | 20,755 | 2017 | (46) |
| **Chicken (*Gallus gallus*)** | Whole-genome shotgun/BAC and other clones (6.6×) | 1.05Gb | 14,923 | 2004 | (47) |

**Table 3.** Genome sequence drafts for the water (Asiatic) buffalo (*Bubalus bubalis*) which are available in the public biological databases

| **Country** | **Submitter (breed)** | **Release year** | **GenBank assembly accession# (publication)** | **Sequencing technology** | **Genome length** | **Assembler** | **Assembly level** | **#Genes** | **Scaffold N50** | **Contig N50** |
|---|---|---|---|---|---|---|---|---|---|---|
| **Egypt** | Egyptian Water Buffalo Genome Consortium (Egyptian buffalo) | 2013 | GCA_002993835.1 (19) | SOLiD | 3.0 Gb | Velvet | Scaffold | NA | 3.5 MB | 14.5 KB |
| **Italy** | University of Maryland, Australia (Mediterranean) | 2013 | GCA_000471725.1 (12) | Illumina GAIIx; Illumina HiSeq; 454 | 2.8 Gb | MaSuRCA | Scaffold | 27,837 | 1.4 MB | 21,9 KB |
| **Bangladesh** | BGI-Shenzhen, China (Bangladesh) | 2012-2019 | GCA_004794615.1 (18) | Illumina HiSeq2000 | 2.7 Gb | SOAPdenovo | Scaffold | 24,613 | 6.9 MB | 25 KB |
| **India** | | 2011 | (41) | WGS, Illumina | 1.56 Gb | (Btau 4.0) | Scaffold | NA | | |
| | Anand Agricultural University, India (Jafarabadi) | 2018 | GCA_000180995.3 (17) | 454; Illumina NextSeq 500 | 3.7 Gb | MaSuRCA | Scaffold | NA | 102 KB | 13,9 KB |
| **Australia** | University of Adelaide, Australia (Mediterranean) | 2018 | GCA_003121395.1 (15) | PacBio | 2.6 Gb | Falcon-Unzip | Chro. | 29,244 | 117.2 MB | 22.4 MB |
| **China** | Guangxi University, China (swamp – and river - buffalo) | 2020 | GWHAAJZ00000000 (14) | PacBio | 2.6 Gb | Wtdbg | Chro. | 20,202 (River buffalo) 19,297 (Swamp buffalo) | 116.1 MB | 3.1 MB |

The data in the table was obtained from the National Centre for Biotechnology Information (NCBI) Assembly database (81).

**Table 4.** Impact of the Axiom Buffalo Genotyping array 90K technology and transcriptome sequencing on buffalo research

| Application | Key findings | Reference |
|---|---|---|
| **Animal Management & Breeding** | SNP discovery, Buffalo Genotyping array 90 Axiom (Affymetrix) | (67) |
| | The ability of genomic selection to reduce generation intervals and improve selection precision (particularly in young bulls), resulting in a very rapid genetic improvement. | (79) |
| **Data Analysis** | AffyPipe has been successfully tested on Buffalo and may be used for any animal genotyped using the Axiom technology. | (70) |
| **Genetic Disorders** | Transverse hemimelia is a hereditary disorder that affects the development of the hindlimbs' distal half. The authors discovered 13 genes that could be implicated in the development of the hindlimbs, implying that the condition is inherited in an oligogenic manner. | (69) |
| **Population Diversity** | The researchers discovered three different gene pools: pure river, pure swamp, and one with genomic mixing between river and swamp buffalo. The historical relationship between populations was also discovered, proving the existence of two separate domestication episodes, one for river buffaloes and the other for swamp buffaloes. | (21) |
| | The study concluded that assessing the degree of LD, phase persistence, and effective population size as a preparatory step before genomic selection and GWASs should be routine. | (59) |
| | The Iranian Buffalo has a higher amount of LD than what is suggested for genome-wide association studies. | (68) |
| **Milk Production** | The discovery of putative genes for reproductive and production traits may be used to search for causative mutations. Four reproductive traits and Six milk production traits were chosen as target phenotypes. | (82) |
| | . A number of interesting genetic regions were found to be connected to daily milk production. | (65) |
| | Buffalo populations structure was investigated using the SNP chip and it was able to distinguish between buffalo from various farms. A GWAS study identified genomic areas on five chromosomes that are thought to be involved in milk production. | (63) |
| | Buffalo milk performance was linked to four SNPs (AX-85148558, AX-85106096, AX-85073877, and AX-85063131) and two genomic areas (3_43 and 14_66). Five genes have been discovered as novel candidate genes for buffalo milk production, including MFSD14A, SLC35A3, PALMD, RGS22, and VPS13B. | (2) |
| | In diverse buffalo breeds, 517 potential genes have been identified to being related to milk production | (3) |
| | The identification of 12 hub genes involved in a variety of milk production pathways. | (66) |
| **Heat tolerance** | There were 753 genes and 16 microRNAs that were differentially expressed in heat-tolerant and non-heat-tolerant Buffalos, according to the study. | (58) |